\documentclass[aps,pre,twocolumn,superscriptaddress,showpacs]{revtex4-1}
\usepackage{amsmath}
\usepackage{amssymb}
\usepackage{tikz}
\usepackage{graphicx}% Include figure files
\usepackage{dcolumn}% Align table columns on decimal point
\usepackage{bm}% bold math
\usepackage{epsfig}
\usepackage{psfrag}
\newcommand{\refsec}[1]{\mbox{Sec.~\ref{#1}}}

\newcommand{\T}{${\mathcal T}\,$}
\newcommand{\Ti}{${\mathcal T}$}

\begin{document}

\title{\bf Non-universality in the spectral properties of time-reversal invariant microwave networks and quantum graphs}
\author{Barbara Dietz}
\email{dietz@ifpan.edu.pl}
\altaffiliation{Present address: School of Physical Science and Technology, and Key Laboratory for Magnetism and Magnetic Materials of MOE, Lanzhou University, Lanzhou, Gansu 730000, China}
\affiliation{Institute of Physics, Polish Academy of Sciences, Al. Lotnik\'ow 32/46, 02-668 Warszawa, Poland}
\author{Vitalii Yunko}
\affiliation{Institute of Physics, Polish Academy of Sciences, Al. Lotnik\'ow 32/46, 02-668 Warszawa, Poland}
\author{Ma{\l}gorzata Bia{\l}ous}
\affiliation{Institute of Physics, Polish Academy of Sciences, Al. Lotnik\'ow 32/46, 02-668 Warszawa, Poland}
\author{Szymon Bauch}
\affiliation{Institute of Physics, Polish Academy of Sciences, Al. Lotnik\'ow 32/46, 02-668 Warszawa, Poland}
\author{Micha{\l} {\L}awniczak}
\email{lawni@ifpan.edu.pl}
\affiliation{Institute of Physics, Polish Academy of Sciences, Al. Lotnik\'ow 32/46, 02-668 Warszawa, Poland}
\author{Leszek Sirko}
\email{sirko@ifpan.edu.pl}
\affiliation{Institute of Physics, Polish Academy of Sciences, Al. Lotnik\'ow 32/46, 02-668 Warszawa, Poland}

\date{\today}

\bigskip

\begin{abstract}
We present experimental and numerical results for the long-range fluctuation properties in the spectra of quantum graphs with chaotic classical dynamics and preserved time-reversal invariance. Such systems are generally believed to provide an ideal basis for the \emph{experimental} study of problems originating from the field of quantum chaos and random matrix theory. Our objective is to demonstrate that this is true only for short-range fluctuation properties in the spectra, whereas the observation of deviations in the long-range fluctuations is typical for quantum graphs. This may be attributed to the unavoidable occurrence of short periodic orbits, which explore only the individual bonds forming a graph and thus do not sense the chaoticity of its dynamics. In order to corroborate our supposition, we performed numerous experimental and corresponding numerical studies of long-range fluctuations in terms of the number variance and the power spectrum. Furthermore, we evaluated length spectra and compared them to semiclassical ones obtained from the exact trace formula for quantum graphs.
\end{abstract}

\bigskip
\maketitle

\section{Introduction\label{Intro}} 

	Quantum graphs~\cite{Kottos1997,Kottos1999,Pakonski2001,Texier2001}, networks of bonds connected at vertices, have been used extensively for the experimental and the theoretical study of closed and open quantum systems, of which the corresponding classical dynamics is chaotic. They were introduced by Linus Pauling to model organic molecules \cite{Pauling1936} and since then used for the modeling of a large variety of systems, a few examples being quantum wires~\cite{Sanchez1988}, optical waveguides~\cite{Mittra1971} and mesoscopic quantum systems~\cite{Kowal1990,Imry1996}. In Ref.~\cite{Gnutzmann2004} the fluctuation properties in the eigenvalue spectra of closed graphs with incommensurable bond lengths were proven rigourously to be described by the Gaussian ensembles (GEs) of random matrix theory (RMT)~\cite{Mehta1990}, in accordance with the Bohigas-Gianonni-Schmit (BGS) conjecture~\cite{Bohigas1984,Guhr1998,Haake2001}. Moreover, graphs have the particular property that the semiclassical trace formula for their spectral density in terms of a sum over the associated periodic orbits is exact~\cite{Keating1991}. Also the correlation functions of scattering matrices describing chaotic scattering on open graphs were shown to coincide with the corresponding RMT results~\cite{Verbaarschot1985,Pluhar2013,Pluhar2013a,Pluhar2014,Fyodorov2005}. Therefore, we may expect that the fluctuation properties in the spectra of classically chaotic quantum graphs with time-reversal (\Ti) invariance, where \Ti$^2=1$, and with violated \T invariance coincide with those of random matrices from the Gaussian orthogonal ensemble (GOE) and the Gaussian unitary ensemble (GUE), respectively, for the case \Ti$^2=1$. This has been shown numerically already in Ref.~\cite{Kottos1999} and also experimentally using microwave networks composed of coaxial cables coupled by junctions at the vertices, which simulate quantum graphs with a chaotic dynamics~\cite{Hul2004,Lawniczak2010}. In the experiments~\cite{Lawniczak2008,Lawniczak2011,Hul2012,Lawniczak2014,Allgaier2014} evidence was provided based on the nearest-neighbor spacing distribution (NNSD). Only recently, microwave networks were succesfully used to realize a system with an antiunitary symmetry \Ti and \Ti$^2=-1$. In these experiments~\cite{Rehemanjiang2016} both the NNSD as a measure for short-range spectral fluctuations and the spectral rigidity $\Delta_3$, also known as Dyson-Mehta statistics~\cite{Mehta1990}, as a measure for long-range spectral fluctuations were shown to coincide with those of random matrices from the Gaussian sympletic ensemble (GSE). 

The $\Delta_3$ statistics has also been investigated in microwave networks simulating \Ti-invariant graphs~\cite{Hul2004,Lawniczak2016} and recently for graphs in which \T invariance was violated~\cite{Bialous2016}. In the latter case, good agreement was found with an extended RMT model applicable to incomplete spectra, whereas in the \Ti-invariant case the deviations observed in the statistical measures for long-range spectral fluctuations could not be fully explained. The objective of this article is an understanding of the discrepancies between the RMT predictions for incomplete and complete spectra and the experimental and the corresponding numerical results, respectively, on the basis of short periodic orbits~\cite{Schanz2003,Gnutzmann2013}. For this, we analyzed long-range spectral fluctuations in the spectra using the average power spectrum~\cite{Relano2002,Faleiro2004,Molina2007} which is directly related to the spectral form factor and the number variance~\cite{Mehta1990}. They were demonstrated to provide powerful tools for the identification of the effects that lead to deviations from the  GOE predictions~\cite{Molina2007}. 

A crucial requirement for the agreement of the fluctuation properties in the eigenvalue spectra of quantum systems with a classically chaotic dynamics with those of the eigenvalues of random matrices from the conventional GEs is the completeness of the spectra~\cite{Bohigas1983,Bohigas1984}. Accordingly, the experimental study of the spectral fluctuation properties might be tedious, especially when using microwave networks in which absorption of microwave power is unavoidable. In order to overcome the problems of absorption and to attain complete sequences of several hundreds of eigenvalues, e.g., in experiments with flat, cylindrical microwave resonators simulating quantum billiards~\cite{Stoeckmann1990,Sridhar1991,Graef1992,So1995,Sirko1997} the measurements, actually, had to be performed with resonators, that were superconducting at liquid-helium temperature~\cite{Dietz2015}. In experiments with microwave networks generally not all, but close to 100\% of the eigenvalues can be found by proceeding as described, e.g., in the present article. 

Missing levels or non-universal contributions, like from the shortest periodic orbits~\cite{Berry1985,Sieber1993}, lead to especially large deviations from the RMT predictions for long-range spectral fluctuation properties. The number variance~\cite{Mehta1990} and the power spectrum~\cite{Relano2002,Faleiro2004,Molina2007}, are particularly suited for the discrimination between deviations caused by missing levels and by non-universal effects. The former corresponds to a two-level correlation function whereas the latter is given in terms of the spectral form factor, i.e., the Fourier transform of the two-level cluster function~\cite{Mehta1990}. Deviations from RMT predictions become visible in the number variance at a few level spacings and, accordingly, in the power spectrum at short times. In order to identify their origin and to ensure that they are intrinsic and not just of experimental nature we, in addition, performed numerical simulations for the quantum graphs corresponding to the microwave networks used in the experiments. Accordingly, we had a larger set of eigenfrequencies at hand than experimentally achievable and thereby were able to unambigiously demonstrate that the deviations from RMT predictions observed in the experiments and the numerical simulations arise due to the presence of short periodic orbits that are localized on the individual bonds by reflection at the vertices terminating them. For this we computed the exact trace formula~\cite{Kottos1999}. Like in Ref.~\cite{Hul2004} we compared it to length spectra obtained from the eigenfrequencies of microwave networks with preserved \T invariance. Furthermore, we computed length spectra using sequences of 2000 numerically obtained eigenfrequencies in order to identify non-universal periodic orbits.     

	In~\refsec{Theory} we will briefly review the salient properties of microwave networks and quantum graphs. Then we will introduce the experimental setup in~\refsec{Exp} and finally present the experimental and numerical results in~\refsec{Anal} and the comparison of their length spectra with those computed from the trace formula for quantum graphs in~\refsec{Semicl}. Furthermore, in~\refsec{Miss} we compare the statistical measures obtained for the experimentally and the numerically determined eigenvalues to results derived for incomplete spectra based on RMT. The results are then summarized in the conlusions.

	\section{Microwave networks and quantum graphs\label{Theory}} 
\begin{figure}[h!]
	\includegraphics[width=\linewidth]{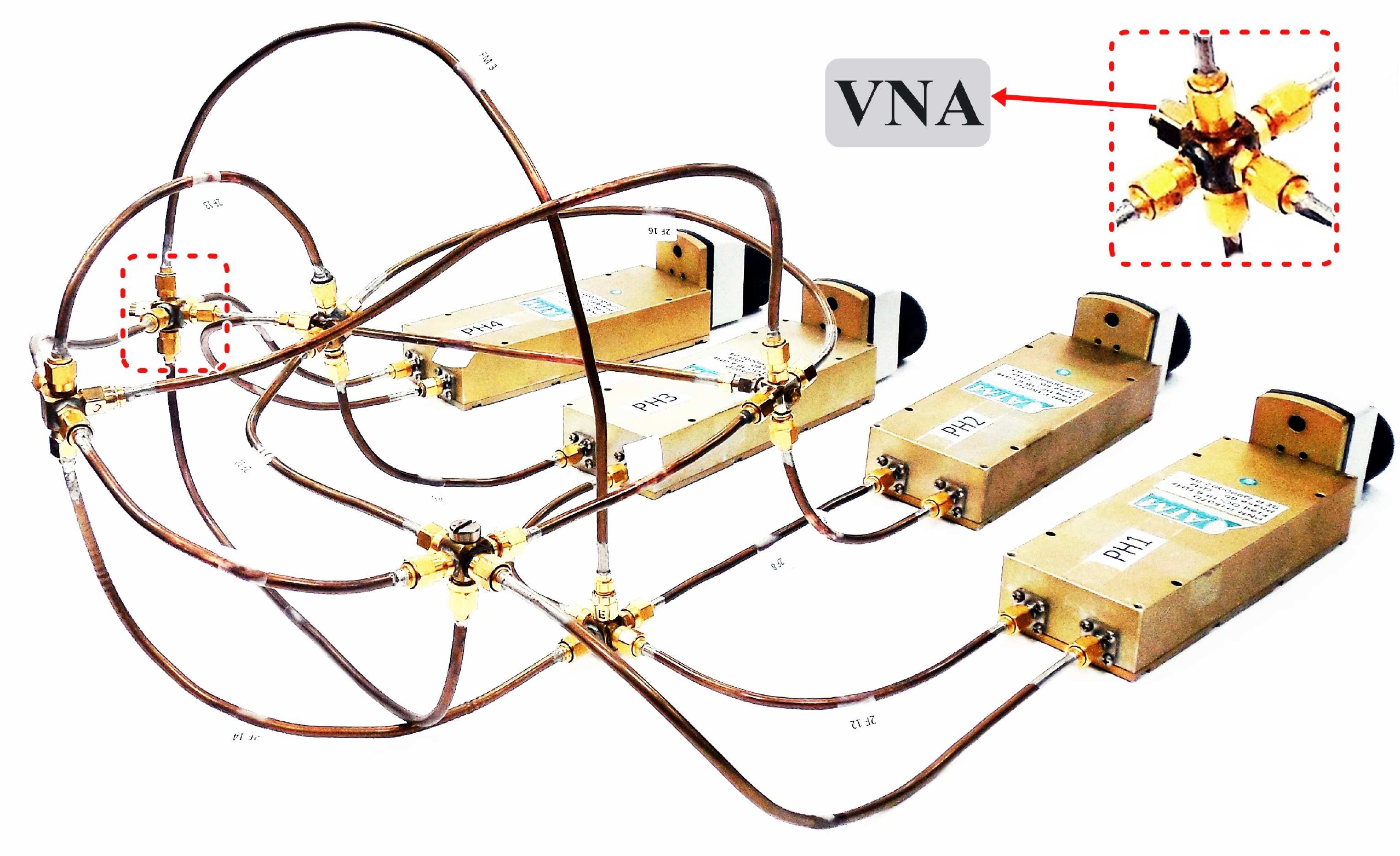}
	\caption{
	Photograph of a microwave network consisting of 6 vertices that were all connected with each other. An ensemble of 30 such networks was created by changing the lengths of four bonds using the phase shifters visible in the background. For the measurement of the scattering matrix, the vector network analyzer (VNA) was coupled to the network via a HP 85133-616 flexible microwave cable as indicated in the inset.} 
	\label{Fig1}
	\end{figure}

	Microwave networks are constructed from coaxial microwave cables, that are coupled by junctions at the vertices. Figure~\ref{Fig1} shows a photograph of a network used in the experiments presented in this article. It consists of 6 junctions, that are all connected with each other by coaxial cables. Thus they simulate fully connected quantum graphs. 
A coaxial cable is composed of an inner and a concentric outer conductor. The space between them is filled with some homogeneous material. Below the cut-off frequency for the first transverse electric (TE$_{11}$) mode only the fundamental transverse electromagnetic (TEM) mode can propagate between the conductors. The geometry of a network of such cables is defined by the connectivity matrix $C_{ij}$, which equals zero if vertices $i$ and $j$ are not connected and unity otherwise, and by the lengths $L_{ij}$ of the coaxial cables connecting the latter. The one-dimensional wave equation of the so-called Lecher waves propagating between the inner and outer conductor along such a coax cable is given in terms of the difference $U_{ij}(x)$ between the potentials at the conductors' surfaces,
\begin{equation}
\frac{{\rm d}^2}{{\rm d}x^2}U_{ij}(x)+\frac{\omega^2\epsilon}{c^2}U_{ij}(x)=0\, ,\, i<j.
\label{WE}
\end{equation}
Here, the coordinate $x$ varies along the coaxial cable from $x=0$ at vertex $i$ to $x=L_{ij}$ at vertex $j$, $\epsilon$ is the dielectric constant of the medium, $\omega =2\pi\nu$ is the angular frequency with $\nu$ the microwave frequency and $c$ is the velocity of light. Equation~(\ref{WE}) is also called telegraph equation. It is applicable to lossless coaxial cables. At each pair of connected vertices $i$ and $j$ the potential difference $U_{ij}(x)$ obeys the continuity equation
\begin{equation}
U_{ij}(x=0)=\phi_i,\, U_{ij}(x=L_{ij})=\phi_j,\, i<j,  
\label{BC1}
\end{equation}
that is, all waves leaving or entering a vertex take there the same value. Furthermore, the current is conserved at each vertex,
\begin{equation}
-\sum_{j<i}C_{ij}\frac{\rm d}{{\rm d}x}U_{ji}(x=L_{ij})+\sum_{j>i}C_{ij}\frac{\rm d}{{\rm d}x}U_{ij}(x=0)=0.
\label{BC2}
\end{equation}
The wave equation~(\ref{WE}) together with the properties Eqs.~(\ref{BC1}) and~(\ref{BC2}) is valid for ideal coaxial cables with vanishing Ohmic resistance. It is mathematically identical with that of a quantum graph with Neumann boundary condition at the vertices and with lengths of the bonds chosen as $L^{opt}_{ij}=\sqrt{\epsilon}L_{ij}$~\cite{Kottos1999,Texier2001}. The corresponding Schr\"odinger equation is obtained by replacing in Eq.~(\ref{WE}) the term $\sqrt{\epsilon}\omega /c$ by the wavenumber $k$. The eigenwavenumbers of the graph are determined by solving the equation~\cite{Kottos1999}
\begin{equation}
\det h_{ij}(k)=0
\end{equation}
with 
\begin{equation}
h_{ij}(k)=\left\{{\begin{array}{cc}
        -\sum_{m\ne i}C_{im}\cot\left(kL_{im}\right)&, i=j\\
        C_{ij}\left(\sin(kL_{ij})\right)^{-1}&, i\ne j\\
        \end{array}}
  \right.\, .
\end{equation}
The spectral density $\rho(k)$ of a quantum graph may be expressed in terms of purely classical quantities. Separating it into its smooth part $\bar\rho(k)$ and its fluctuating part $\rho^{fluc}(k)$, the former is given by Weyl's formula,
\begin{equation}
\bar\rho (k)=\frac{\mathcal{L}}{\pi},
\label{Weyl}
\end{equation}
where $\mathcal{L}$ denotes the total length of the graph, the latter by an exact trace formula 
\begin{equation}
\label{trace}
\rho^{fluc}(k)=\frac{1}{\pi}\sum_{p\in\mathcal{P}_n}\frac{l_{p}\cos\left(r\left[kl_p+\pi\mu_p\right]\right)}{e^{r(n_p\gamma_p/2)}}
\end{equation}
with 
\begin{equation}
e^{-n_p\gamma_p/2}=\prod_{s=1}^{\mu_p}\left\vert\left(1-\frac{2}{v_s}\right)\right\vert\prod_{s=1}^{n_p-\mu_p}\left\vert\frac{2}{v_s}\right\vert\, .
\end{equation}
The sum is over the set of primitive periodic orbits $p\in\mathcal{P}_n$, coded by a sequence of $n_p$ vertices with $n=rn_p$ denoting the period after $r$ repititions of it. The length of the primitive periodic orbit, which is given by the sum over the lengths of the bonds that are passed during one loop of it, is denoted by $l_p$. Furthermore, $\mu_p$ gives the number of vertices with $v_i\geq 2$, where backscattering occurs. Here, $v_i$ corresponds to the valency of vertex $i$, that is, the number of bonds coupled to it. In the numerical simulations the valency of each of the six vertices equaled five, and also the lengths of the bonds were chosen equal to the optical lengths of the coaxial cables in the 30 microwave networks used in the experiments. 
    
\section{Experimental setup\label{Exp}}
The microwave networks were constructed from 6 vertices that were all connected with each other via coaxial cables (SMA-RG402) that consisted of an inner conductor of radius $r_1=0.05$~cm and an outer one with inner radius $r_2=0.15$~cm. The space between them was filled with Teflon of which the dielectric constant was determined experimentally to $\varepsilon\simeq 2.06$. Accordingly,  the cut-off frequency  $\nu_{c}=\frac{c}{\pi (r_1+r_2)\sqrt{\varepsilon}}$~\cite{Jones,Savytskyy2001} of the TE$_{11}$ mode was at around 33~GHz, which is well above the range of frequencies used in the experiments. The lengths of the cables were chosen such that the microwave networks simulated quantum graphs with chaotic dynamics. An ensemble of 30 different networks with the same total optical length $\mathcal{L} = 7.04\pm0.02$~m was generated by varying the lengths of four bonds of lengths $L_i\simeq 40-65$~cm with phase shifters (see Fig.~\ref{Fig1}) in steps of $\pm\, 0.42$~cm. Here, $\mathcal{L}$ was estimated by measuring the optical length of each coaxial cable including their extensions across the junctions separately.

\begin{figure}[h!]
	\includegraphics[width=\linewidth]{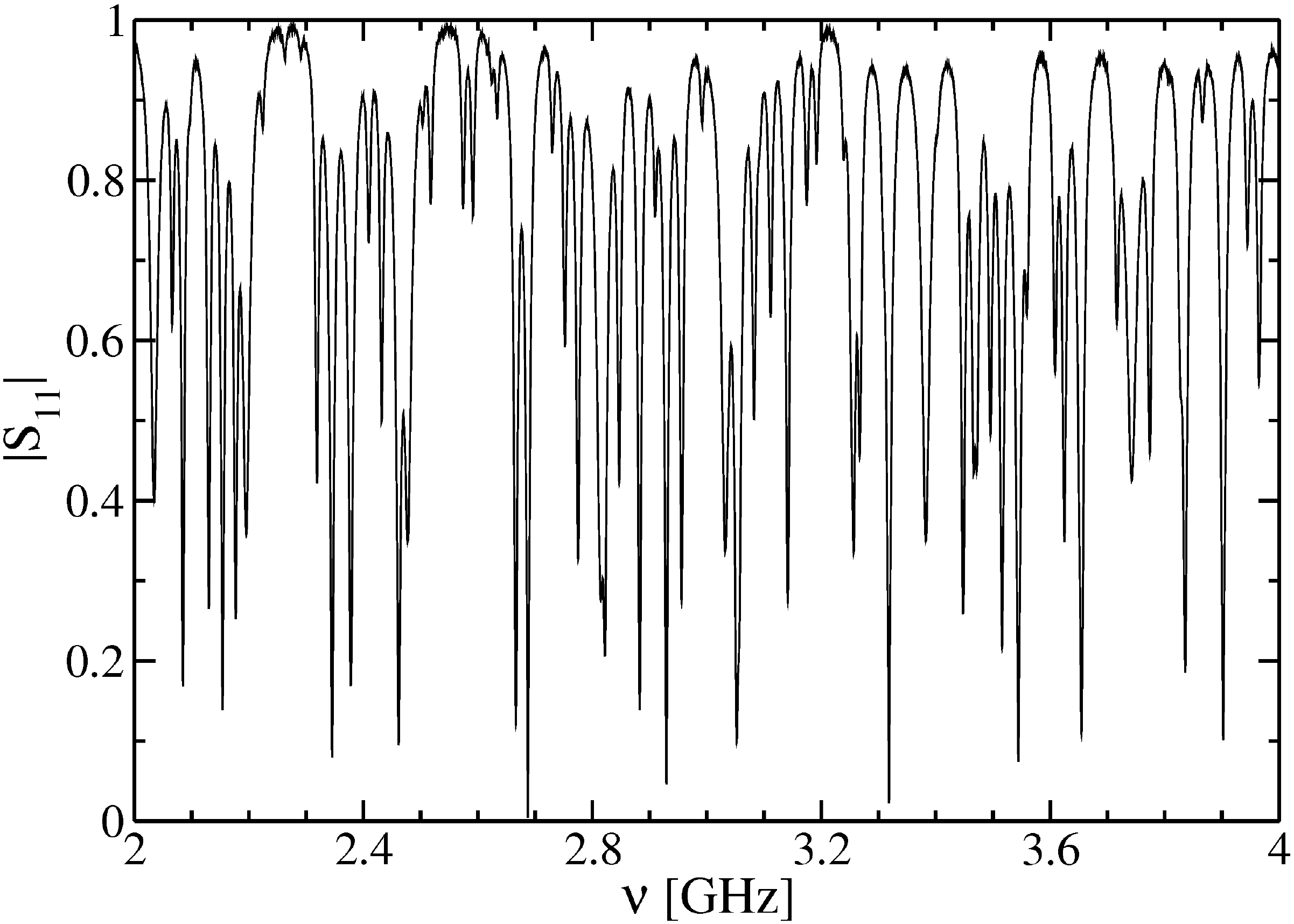}
	\caption{
	A reflection spectrum in the frequency range from 2-4~GHz. The overlap between neighboring resonances was sufficiently weak below 6~GHz, so that we were able to determine about 96-97~\% of the eigenfrequencies for each of the corresponding quantum graphs.
	}\label{Fig2}
	\end{figure}

\begin{figure}[h!]
	\includegraphics[width=\linewidth]{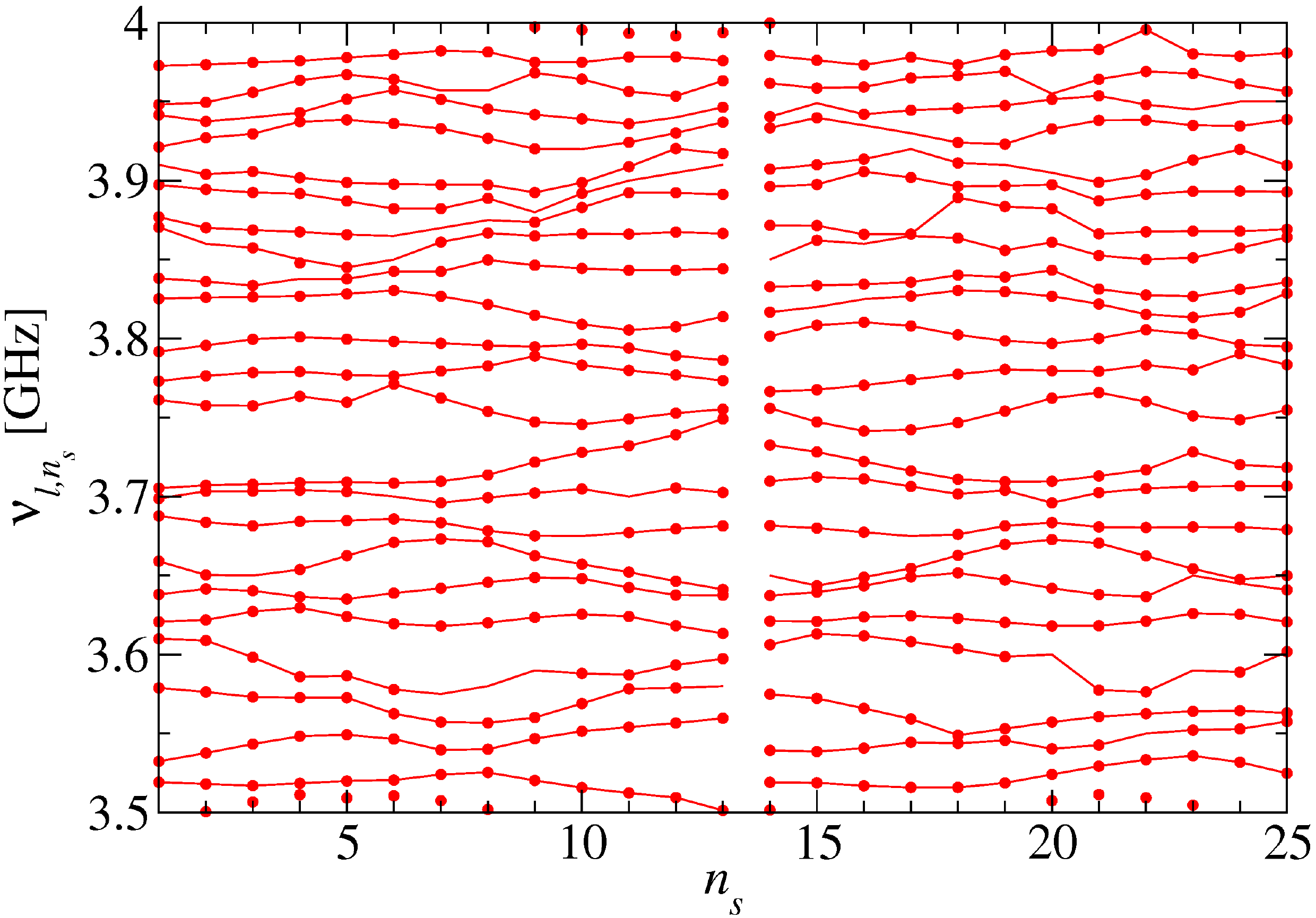}
	\caption{
	Sequences of the eigenfrequencies of the 30 graphs labelled by the step number $n_s$ in the frequency range from 3.5-4~GHz. The graph with label $n_s+1$ was obtained from that with label $n_s$ by increasing the lengths of two bonds by 0.42~cm and at the same time decreasing that of two other ones by the same amount so the total length was fixed. At $n_s=13$ and $n_s=25$ for one of these bonds increase was changed into decrease, and vice versa for another one. The resulting eigenfrequency dynamics was used to decide whether a hump in a resonance corresponds to a spurious or a genuine eigenfrequency. Missing red dots at a given $n$ correspond to non-detectable eigenfrequencies.
	}\label{Fig3}
	\end{figure}

\begin{figure}[h!]
    \begin{tikzpicture}
        \node[anchor=south west,inner sep=0] (image) at (0,0){\includegraphics[width=1.0\linewidth]{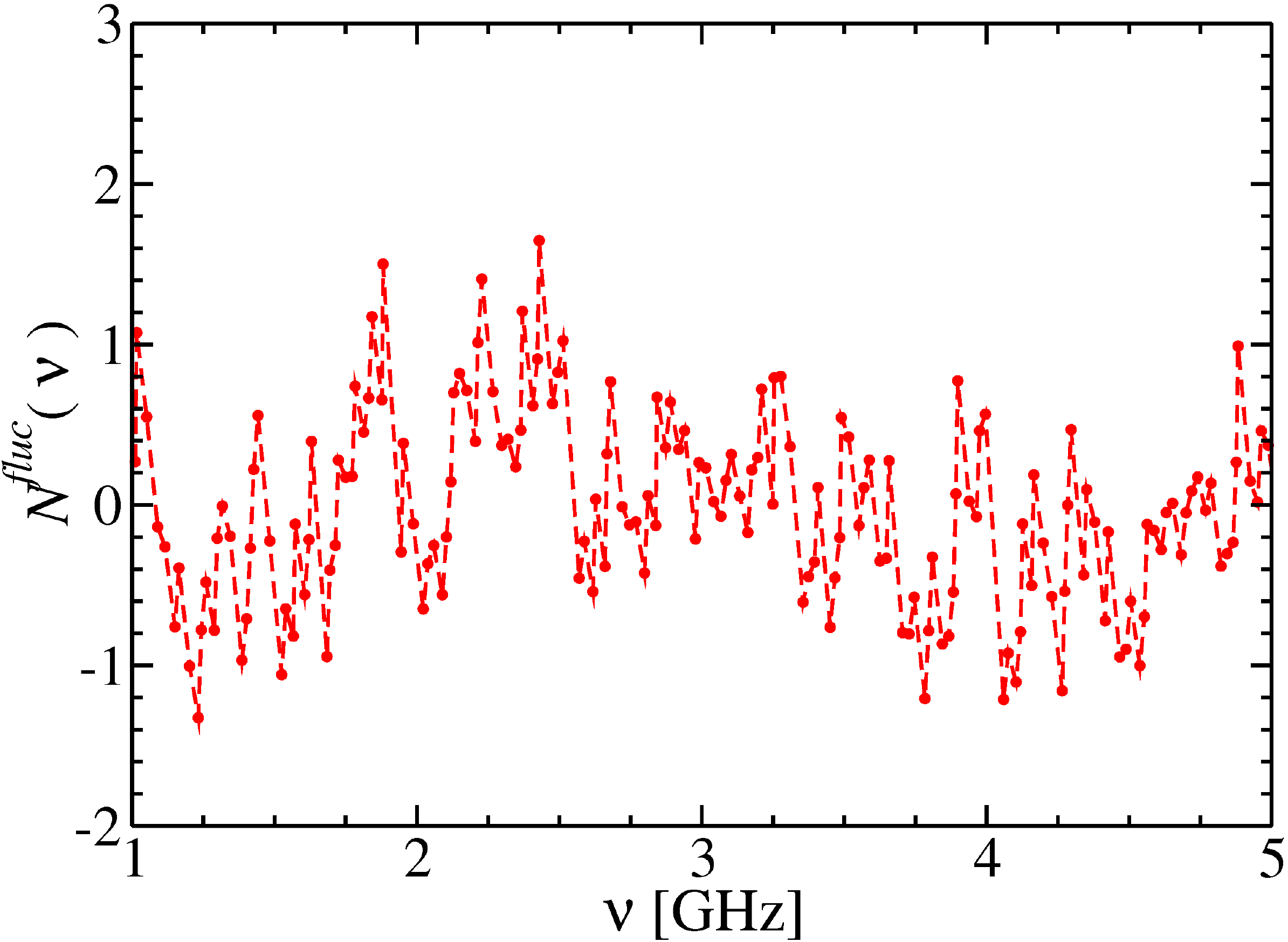}};
\begin{scope}[x={(image.south east)},y={(image.north west)}]
                 \node[anchor=south west,inner sep=0] (image) at (0.64,0.63) {\includegraphics[width=0.32\linewidth]{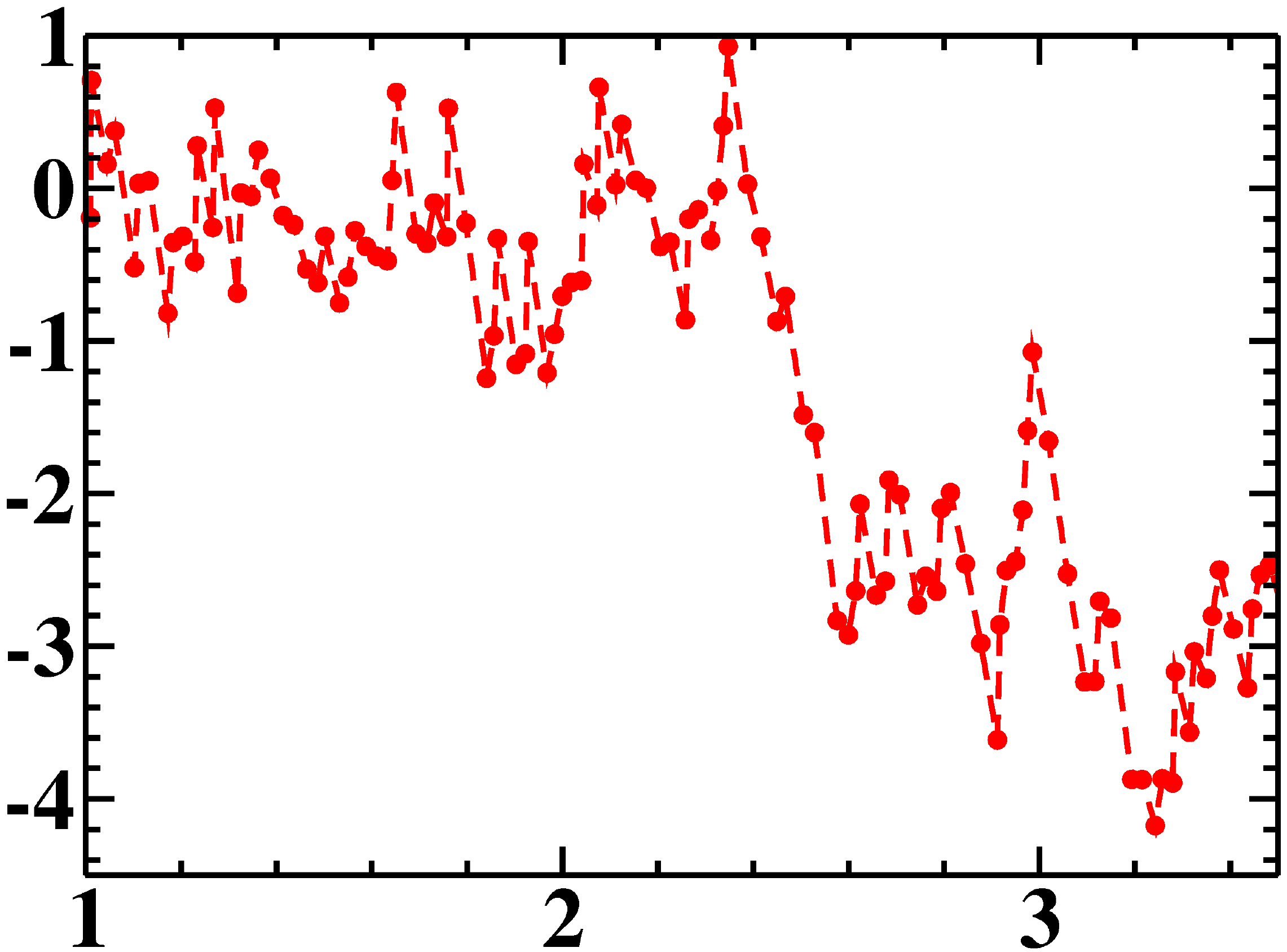}};
        \end{scope}
    \end{tikzpicture}
	\caption{
	Fluctuating part $N^{fluc}(\nu)$ of the integrated spectral density. Each dot corresponds to one eigenfrequency. A missing eigenfrequency results in a jump of the local average of $N^{fluc}(\nu)$ by more than one. This is illustrated in the inset, where two eigenfrequencies are missing around 2.4-2.5 GHz.
	}\label{Fig4}
	\end{figure}

For the measurement of the scattering matrix element $S_{11}$ an Agilent E8364B microwave vector network analyzer (VNA) was connected to one six-arm vertex of the network via a HP 85133-616 flexible microwave cable; see inset in Fig.~\ref{Fig1}. Figure~\ref{Fig2} shows a measured reflection spectrum in the frequency range from 2-4~GHz. The positions of the resonances yield the eigenfrequencies and thus the eigenvalues of the corresponding quantum graph. Ohmic losses in the coaxial cables lead to a broadening of the resonances. Below 6~GHz thereby arising overlaps between the resonances were sufficiently weak so that we were able to determine nearly all eigenfrequencies by fitting Lorentzians to them. In order to decide whether a hump in a broad resonance corresponds to a genuine eigenvalue we used two very efficient methods. First, we plotted the level sequences obtained for the 30 microwave networks, which were generated by stepwise changing the lengths of four bonds while keeping the total length fixed, versus the number of steps thus yielding continuously varying eigenfrequencies (see Fig.~\ref{Fig3}), i.e., a level dynamics~\cite{Haake2001}. Here, the lengths of two of them were increased, and those of the other two were decreased by the same amount. After 13 and 25 steps, respectively, in one bond the increase was changed to decrease, and vice versa in another one, to keep the total length of the network fixed. In order to indicate the level dynamics we connected eigenfrequencies (red dots) by red lines. Then missing levels become visible as gaps in the level dynamics. Second, we looked at the fluctuating part of the integrated spectral density $N^{fluc}(\nu_i)$, that is, the difference of the number of identified eigenfrequencies $N(\nu_i)=i$ below $\nu_i$ for ordered frequencies $\nu_{1}\leq\nu_{2}\leq\dots$ and the number predicted by Weyl's formula~\cite{Kottos1999} given in Eq.~(\ref{Weyl}). One example is shown in Fig.~\ref{Fig4}. At a missing or spurious eigenfrequency the local average of $N^{fluc}(\nu_i)$ exhibits jumps by more than 1. This is illustrated in the inset of  Fig.~\ref{Fig4}. There two eigenfrequencies are missing around 2.4-2.5~GHz. Approximately 210 eigenfrequencies could be identified for each network. A comparison with Weyl's formula yielded that about 3-4\% of the eigenvalues were missing after application of the above described procedure. Actually, Figs.~\ref{Fig3} and~\ref{Fig4} show parts of the final result. Still missing eigenfrequencies might arise due to the overlap of a too close encounter of two levels or because of a vanishing electric field strength at the position of a resonance, resulting in a vanishing or too small amplitude of it to be detectable.   

	\section{Analysis of the experimental and numerical data\label{Anal}}
	For the analysis of the spectral properties of the microwave networks, first, the eigenfrequencies $\nu_i$ need to be unfolded in order to eliminate system specific properties. This was done with the help of Weyl's law given in Eq.~(\ref{Weyl}). Accordingly, the unfolded eigenvalues are obtained from the ordered eigenfrequencies as $\epsilon_i=(2\nu_i/c)\mathcal{L}$. 

	In order to investigate short-range spectral fluctuations we analysed the commonly used NNSD, that is, the distribution of the spacings between adjacent eigenvalues, $s_i=\epsilon_{i+1}-\epsilon_i$. For the study of long-range spectral fluctuations we considered the number variance and the average power spectrum~\cite{Mehta1990,Relano2002,Faleiro2004}. The latter is obtained from the Fourier spectrum of the deviation of the $q$th-nearest neighbor spacing from its mean value $q$, $\delta_q=\epsilon_{i+q}-\epsilon_i-q$, 
	\begin{equation}
	\label{Eq.1}
	S(\tilde k)=\left\vert\tilde{\delta}_{\tilde k}\right\vert^2=\left\vert\frac{1}{\sqrt{N}}\sum_{q=0}^{N-1} \delta_q\exp\left(-2\pi i\tilde kq\right)\right\vert^2
	\end{equation}
with $N$ denoting the number of levels taken into account and $\tilde k\leq 1$ takes the values $1/N,\, 2/N,\,\cdots,\, (N-1)/N$. For $\tilde k\ll 1$ the power spectrum exhibits a power law dependence $\langle S(\tilde k)\rangle\propto \tilde k^{-\alpha}$, where for regular systems $\alpha =2$ and for chaotic ones $\alpha =1$ independently of whether \T invariance is preserved or not~\cite{Relano2002,Faleiro2004,Robnik2005,Salasnich2005,Santhanam2005,Faleiro2006,Relano2008,Bialous2016a}. Recently, the power spectrum was successfully applied to the measured molecular resonances in $^{166}$Er and $^{168}$Er~\cite{Frisch2014,Mur2015} and extended to systems with violated \T invariance~\cite{Bialous2016}. 

\begin{figure}[h!]
    \begin{tikzpicture}
        \node[anchor=south west,inner sep=0] (image) at (0,0){\includegraphics[width=1.0\linewidth]{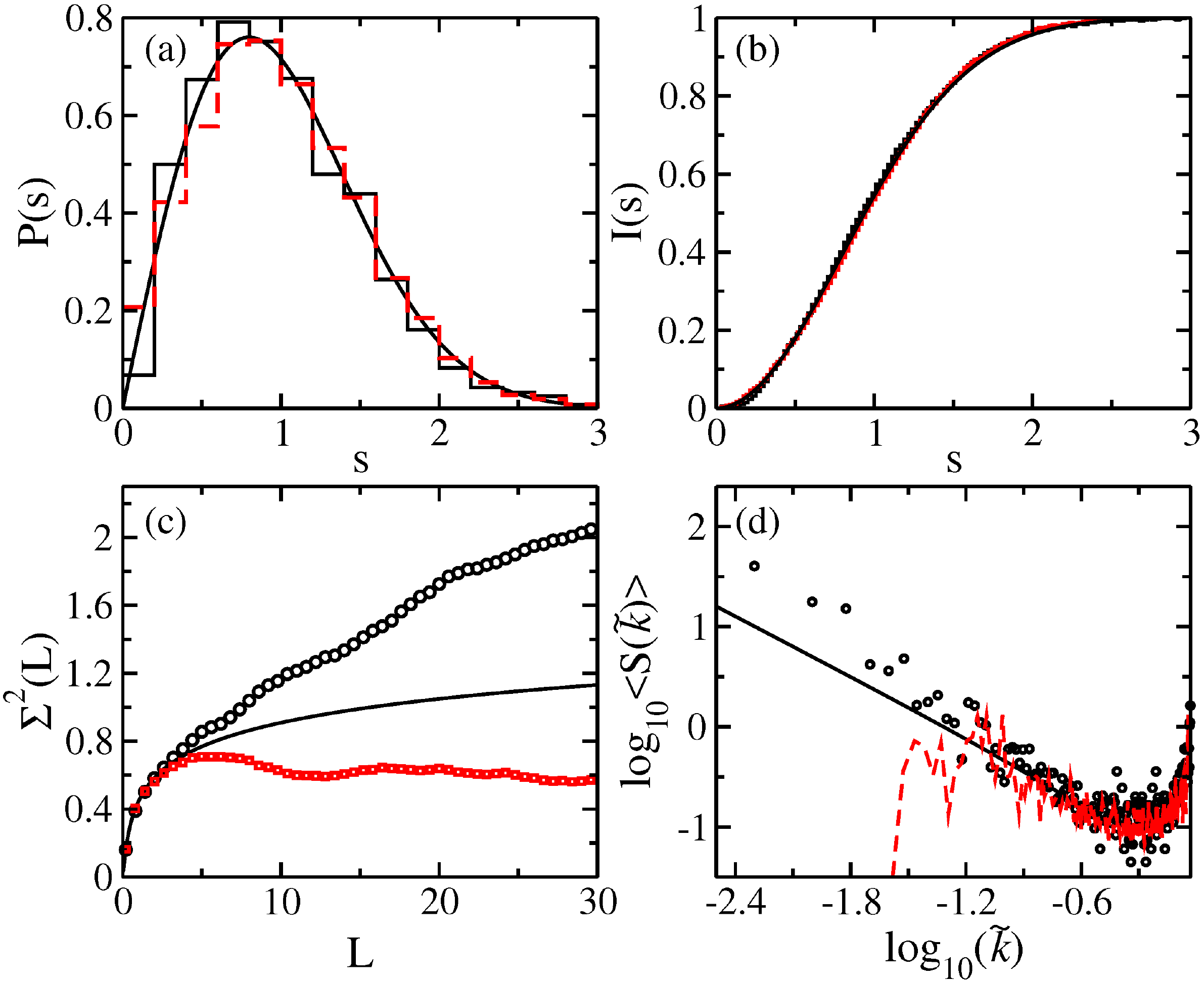}};
\begin{scope}[x={(image.south east)},y={(image.north west)}]
                 \node[anchor=south west,inner sep=0] (image) at (0.74,0.61) {\includegraphics[width=0.23\linewidth]{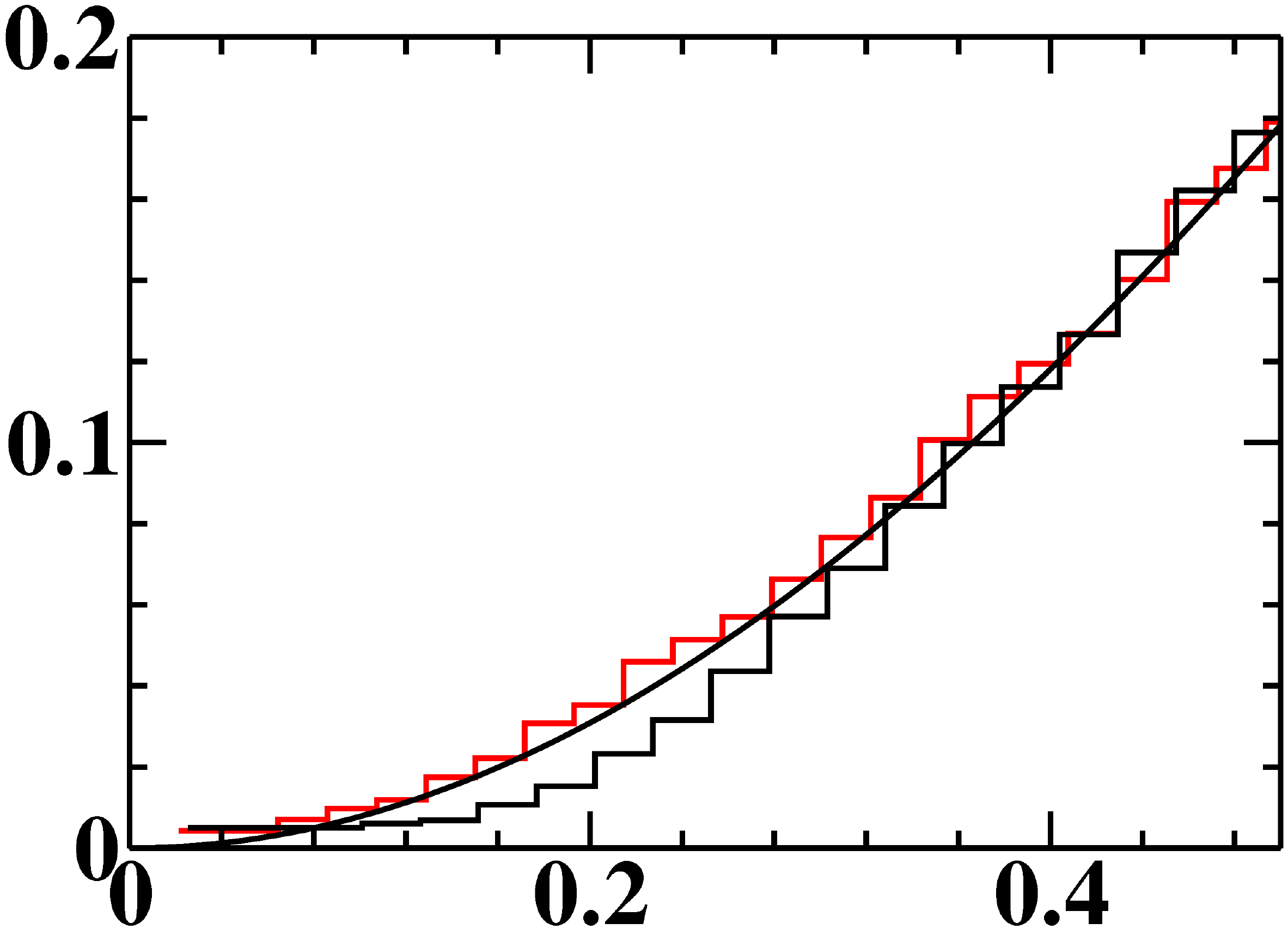}};
        \end{scope}
    \end{tikzpicture}
\caption{(Color online) Spectral properties of the unfolded eigenfrequencies. Panels (a)-(d) show the NNSD, the integrated NNSD, the number variance and the average power spectrum, respectively. The experimental results (black histograms and circles) are compared to the GOE curves (solid black lines) and the corresponding numerical results (red [dark gray] histograms, dashed lines and squares).
}\label{Fig5}
\end{figure}

		Figure~\ref{Fig5} shows in (a) the NNSD $P(s)$, in (b) the integrated NNSD $I(s)$, which was obtained by counting the spacings with values below a given $s$ and dividing by the total number of spacings, in (c) the number variance $\Sigma^2(L)$ and in (d) the power spectrum $\langle S(\tilde k)\rangle$ plotted on a double-log scale for better visibility. The experimental curves (black histograms and circles) were generated by computing the statistical measures for each of the 30 microwave networks and then performing an ensemble average over the results obtained for every third sequence in Fig.~\ref{Fig3} in order to ensure statistical independence of the associated eigenfrequencies. Since the classical dynamics is chaotic and \T invariance is preserved, they are compared to the corresponding GOE results (solid black lines). While in the short-range spectral fluctuations in (a) and (b) only small deviations from the GOE predictions are observed for small spacings, discrepancies are clearly visible for the long-range spectral fluctuations in (c) and (d). Similar results were obtained for the numerically determined eigenvalues of the corresponding quantum graphs (red histograms, dashed lines and squares). Both for the experimentally and the numerically determined eigenvalues the number variance and the power spectrum deviate from the GOE result for $L\gtrsim 3$ and $\tilde k\lesssim -0.9$, respectively. These discrepancies can not be solely attributed to missing levels, because we used complete sequences of up to 2000 eigenvalues for the numerical calculations. In this case the $\Sigma^2$ statistics and the power spectrum saturate below the GOE curve~\cite{Berry1985} whereas the corresponding curves obtained from the experimental data, evolve above the GOE results. Note, that the numerical calculations are performed for closed graphs whereas, due to the connection of the external lead, i.e., the HP85133-616 microwave cable to the VNA and due to Ohmic losses in the coaxial cables, the microwave networks correspond to slightly opened quantum graphs. However, the aim of the present studies was not a level-by-level comparison of numerical and experimental data, but the statistical analysis of their spectral fluctuations and the comparison with RMT predictions. We demonstrate in Sect.~\ref{Miss} that this deviant behavior, actually, may be attributed to a small percentage of missing levels in the experimental level sequences. In the following section we will pursue the supposition that the deviations originate from short periodic orbits.

\section{Trace formula for quantum graphs\label{Semicl}}
\begin{figure}[h!]
\includegraphics[width=\linewidth]{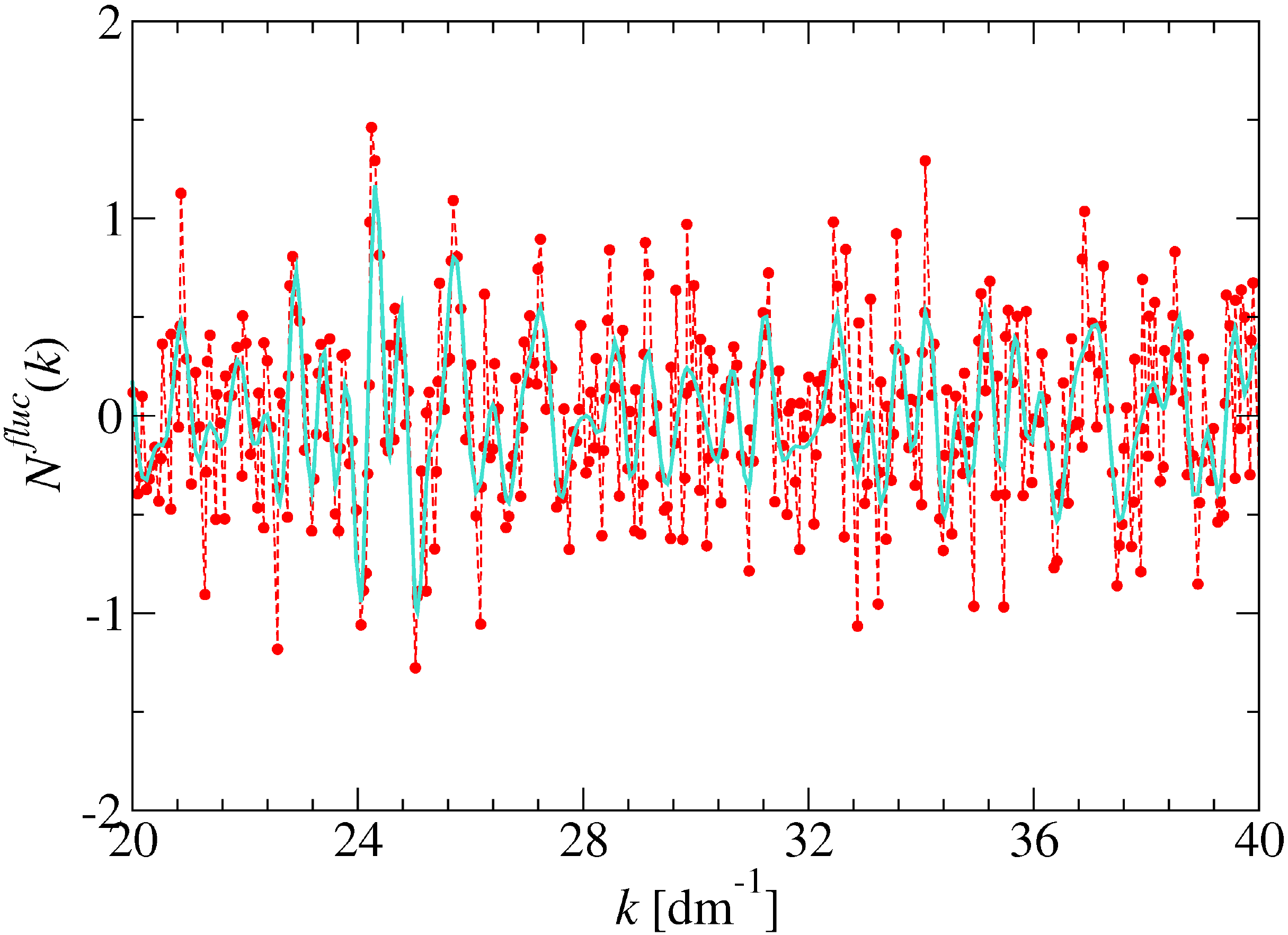}
\caption{(Color online)
Fluctuating part of the integrated spectral density (thin red [dark gray] line), obtained by using 1800 computed eigenvalues of one of the quantum graphs. It exhibits slow oscillations which are well described by the associated integrated semiclassical trace formula Eq.~(\ref{trace}), taking into account only the periodic orbits confined to a bond via reflections at the vertices at its ends (solid turquoise [gray] line). 
}\label{Fig61}
\end{figure}
\begin{figure}[h!]
\includegraphics[width=\linewidth]{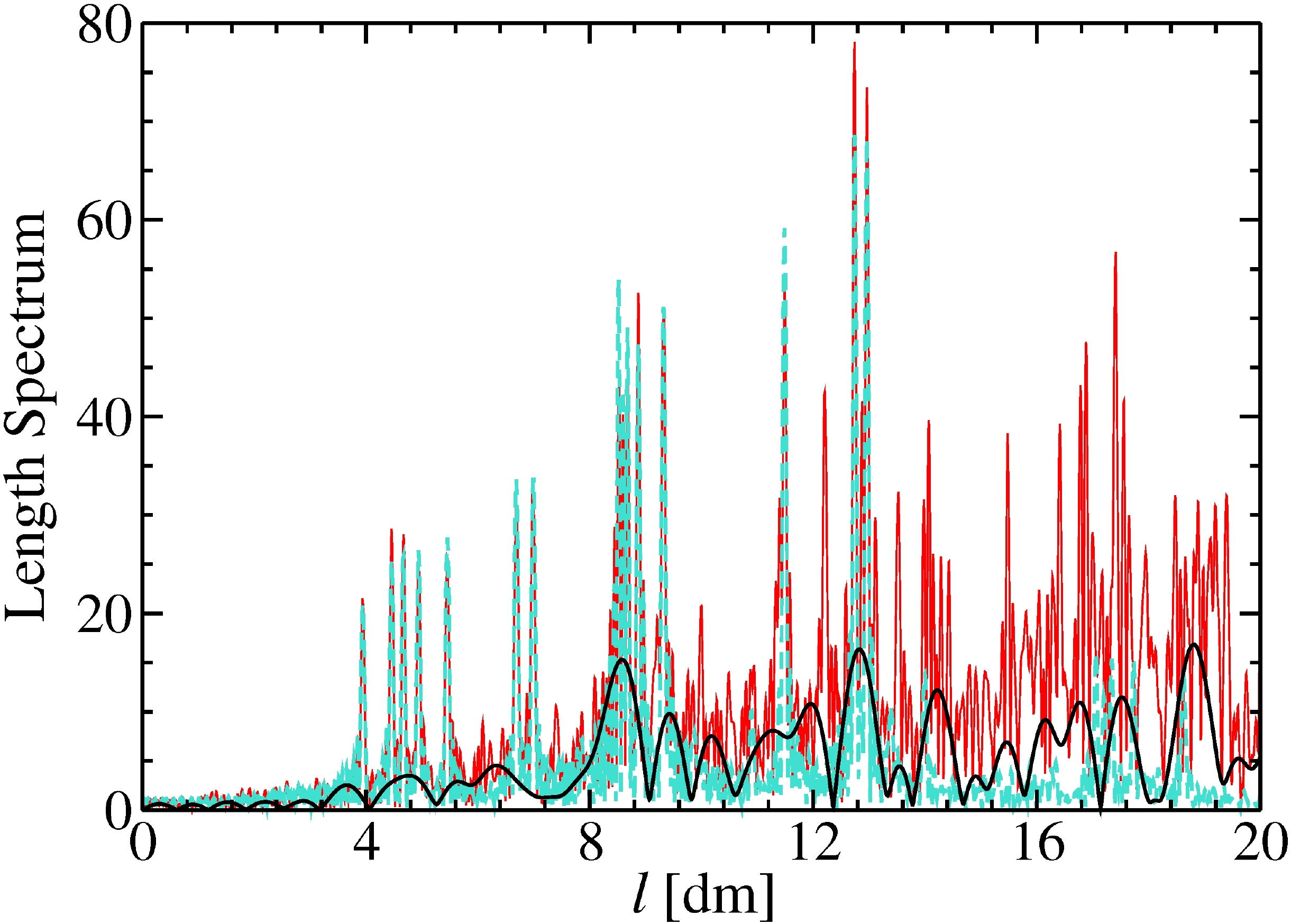}
\caption{(Color online)
Comparison of an experimental (solid black line) and the corresponding numerical (thin red [dark gray] line) length spectrum with the Fourier transform of the semiclassical trace formula Eq.~(\ref{trace}) for the periodic orbits confined to a bond via reflections at the vertices at its ends (dashed turquoise [gray] line) which, accordingly, exhibits peaks at twice the lengths of the bonds. The experimental length spectrum was obtained by using 200 eigenfrequencies of one of the microwave networks. For the numerical one 1800 eigenvalues were taken into account, in order to illustrate the applicability of the trace formula and to identify the peaks corresponding to lengths of periodic orbits confined to one of the 15 bonds.
}\label{Fig6}
\end{figure}
In order to ascertain that the non-universal effects originate from the shortest periodic orbits, which are confined to, respectively, one of the bonds, we computed length spectra using the experimental and the numerical data. A length spectrum is obtained from the Fourier transform of the fluctuating part of the density of eigenwavevectors,   
\begin{equation} 
\vert\tilde\rho (l)\vert =\left\vert\int_0^{k_{max}}dke^{ikl}\rho^{fluc}(k)\right\vert\, .
\label{FFT}
\end{equation}
It is called length spectrum because it exhibits peaks at the lengths of the classical periodic orbits. In order to identify the above-mentioned shortest orbits, we equally calculated the Fourier transform of the trace formula Eq.~(\ref{trace}), which provides the exact semiclassical description of the fluctuating part of the spectral density of quantum graphs. Here, we took into account only those periodic orbits, which are confined to one of the bonds. Accordingly, we restricted the sum in Eq.~(\ref{trace}) to primitive orbits with twice the lengths of a bond, that is, with $l_p=2L_{ij}$, $n_p=2$, $\mu_p=n_p$. Its integral, shown as solid turquoise line in Fig.~\ref{Fig61} for one of the 30 graphs describes the slow oscillations exhibited by the fluctuating part of the numerically obtained integrated spectral density of the associated graph (red dots). The resulting semiclassical length spectrum is shown in Fig.~\ref{Fig6} as turquoise dashed line. The corresponding length spectrum obtained from the 200 experimental eigenfrequencies is shown as solid black line and that obtained from the numerical data as thin red line. Here, we used 1800 computed eigenvalues in order to achieve agreement with the semiclassical length spectrum for the orbits confined to a bond, thereby illustrating that indeed the shortest periodic orbits provide a considerable contribution to the length spectrum. Also the length spectrum deduced from the experimental data shows peaks at the lengths of the shortest periodic orbits. Thus, evidently, backscattering at the joints of the microwave networks and the resultant confinement of waves to one of the bonds is non-negligible. These waves do not experience the joint effects arising from the scattering at all vertices of the quantum graph, that lead to the chaoticity of the underlying dynamics. This explains the deviations of the long-range fluctuations in the spectra of the microwave networks and the quantum graphs from the predicted GOE behavior. 

Note, however, that we didn't observe such deviations in the experiments with microwave networks with violated \T invariance. In these experiments circulators were used, which hinder the backscattering. The discrepancies observed in these experiments between the spectral properties of the microwave networks and the GUE results could be shown unambiguously to be exclusively due to a small percentage of missing levels~\cite{Bialous2016}. We, in fact, also observe similar discrepancies in the present case. As is clearly visible in Fig.~\ref{Fig5}, the curves obtained from the experimental eigenfrequencies for the number variance and the power spectrum lie above the GOE predictions, whereas the numerical ones lie below. In order to ascertain that this different behavior may be attributed to a small percentage of missing levels in the experimental level sequences, we performed similar studies as in Ref.~\cite{Bialous2016}. These are presented in the following section.   

\section{Missing-level statistics\label{Miss}}
The incompleteness of level sequences is a very common problem in experiments involving, e.g., nuclei or molecules~\cite{Liou1972,Zimmermann1988,Agvaanluvsan2003,Agvaanluvsan2003a} thus rendering the study of spectral properties a tedious task. Analytical expressions were derived for the statistical measures describing the fluctuation properties in incomplete spectra based on RMT in Ref.~\cite{Bohigas2004}. The resulting NNSD is given by a sum over the $(n+1)$st nearest-neighbor spacing distributions $P(n,s),\,  n=0,1,2$, with $P(0,s)=P(s)$ which are well approximated by $P(n,s)\simeq\gamma s^\mu e^{-\varkappa s^2}$, with $(n,\mu)=(0,1),\, (1,4),\, (2,8),\cdots$ for the GOE~\cite{Stoffregen1995}. The coefficients $\gamma$ and $\varkappa$ are obtained from the normalization of $P(n,s)$ to unity and the scaling of $s$ to average spacing unity, respectively. As long as the fraction of detected eigenvalues $\varphi$ is close to unity, the NNSD is well approximated by
\begin{equation}
p(s)\simeq P\left(\frac{s}{\varphi}\right)+(1-\varphi)P\left(1,\frac{s}{\varphi}\right)+.... 
\label{abst}
\end{equation}
Furthermore, the number variance and the power spectrum~\cite{Molina2007} are given in terms of the corresponding expressions for complete spectra ($\varphi =1$),
\begin{equation}
\sigma^2(L)=(1-\varphi)L+\varphi^2\Sigma^2\left(\frac{L}{\varphi}\right)
\label{sigma2}
\end{equation}
and 
\begin{eqnarray}
\langle s(\tilde k)\rangle &=&\nonumber
\frac{\varphi}{4\pi^2}\left[\frac{K\left(\varphi\tilde k\right)-1}{\left(\tilde k\right)^2}+\frac{K\left(\varphi\left(1-\tilde k\right)\right)-1}{(1-\tilde k)^2}\right]\\
&+& \frac{1}{4\sin^2(\pi\tilde k)} -\frac{\varphi^2}{12},
\label{noise}
\end{eqnarray}
respectively. Here, $K(\tau)$ is the spectral form factor, where $\tau\leq 1$ in Eq.~(\ref{noise}), so $K(\tau)=2\tau -\tau\log(1+2\tau)$ for the GOE. 

\begin{figure}[h!]
    \begin{tikzpicture}
        \node[anchor=south west,inner sep=0] (image) at (0,0){\includegraphics[width=1.0\linewidth]{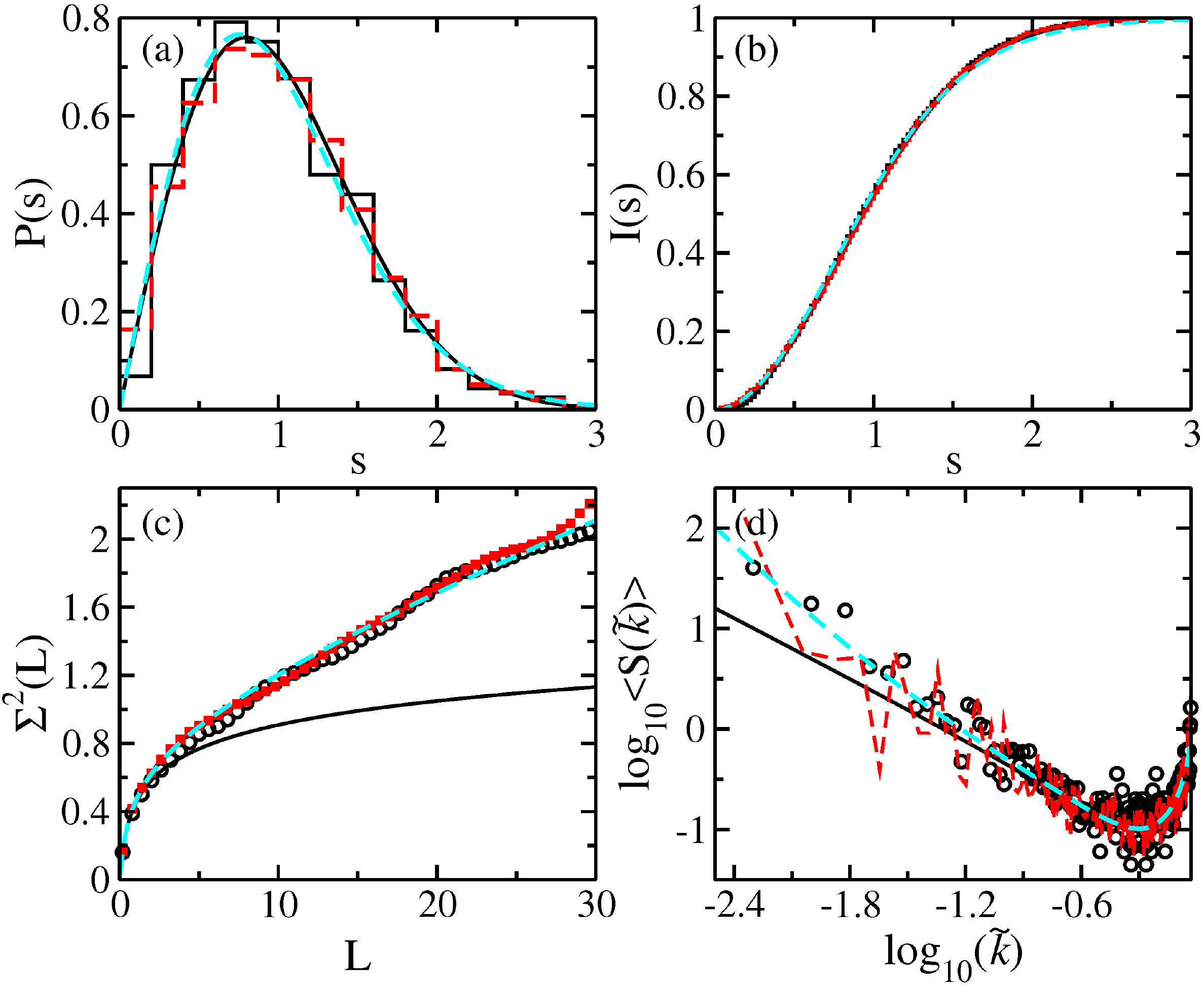}};
\begin{scope}[x={(image.south east)},y={(image.north west)}]
                 \node[anchor=south west,inner sep=0] (image) at (0.74,0.61) {\includegraphics[width=0.23\linewidth]{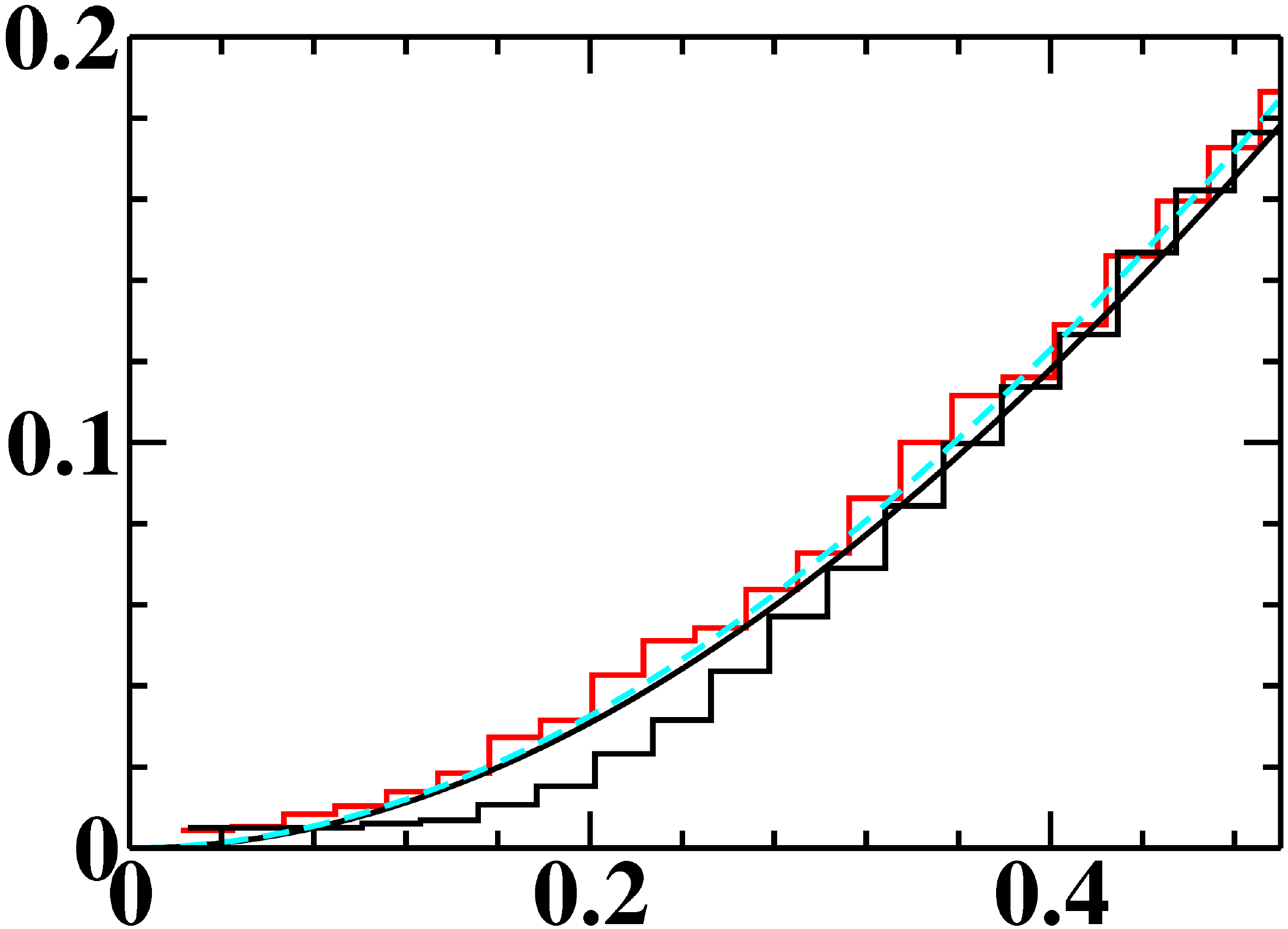}};
        \end{scope}
    \end{tikzpicture}
\caption{
(Color online) Spectral properties of the unfolded eigenfrequencies. Panels (a)-(d) show the NNSD (histogram), the integrated NNSD (circles), number variance (circles) and the average power spectrum (circles), respectively. The experimental results are compared to the GOE curves (solid black lines) and to the theoretical curves Eqs.~(\ref{abst})-(\ref{noise}) for $\varphi =0.965$ (turquoise [gray] dashed lines). The corresponding numerical results (red [dark gray] dashed lines and squares) where obtained by using sequences of 200 eigenvalues and randomly eliminating 3.5~\% of them.
}\label{Fig7}
\end{figure}

In Fig.~\ref{Fig7} the functions Eqs.~(\ref{abst})-(\ref{noise}) are plotted for $\varphi=0.965$ as turquoise dashed lines. This value of $\varphi$ was determined by fitting the analytical results Eqs.~(\ref{sigma2}) and~(\ref{noise}) to the corresponding experimental ones. Furthermore, we evaluated the statistical measures for the numerically determined eigenvalues taking into account a sequence of 200 eigenvalues and randomly eliminating 3.5\% from it. The agreement between the experimental (black histograms and circles) and numerical (red histograms, dashed lines and squares) results is remarkable, also that with the RMT results. However, small differences are visible for the long-range spectral fluctuations in the region where in Fig.~\ref{Fig5} the onset of deviations from the GOE is observed, that is, around $L\approx 3$ in Fig.~\ref{Fig5} (c) and $\log_{10}(\tilde k)\approx -0.9$ in Fig.~\ref{Fig5} (d). There the effect of non-univeral contributions from the shortest periodic orbits becomes perceptible. Like the experimental NNSD the corresponding curve  Eq.~(\ref{abst}) indeed is very close to the GOE results. This feature enabled the assignment of the GOE as the RMT model applicable to the experimental data. The discrepancies between the experimental and RMT curves visible in the integrated NNSD at small spacings (see insets in Figs.~\ref{Fig5} and~\ref{Fig7}) may be attributed to the experimental resolution which impedes the accurate determination of eigenfrequencies if they are to close to each other. Note, that the deviations observed between the statistical measures for long-range spectral fluctuations obtained from the experimental and the numerical data and the RMT predictions taking into account missing levels, are much smaller than those between the numerical, i.e., complete spectra and the GOE result in Fig.~\ref{Fig5}.    

\section{Conclusions\label{Concl}}
We showed experimentally and numerically that deviations from GOE predictions for incomplete and complete spectra observed for long-range spectral fluctuations in the level sequences of quantum graphs, consisting of 6 vertices that are connected with each other via bonds of incommensurable lengths, may be attributed to the occurrence of short periodic orbits confined to individual bonds by backscattering at the vertices terminating them. Such orbits do not sense the chaoticity of the underlying classical dynamics, which arises due to a joint effect of the scattering at all its vertices. In addition, their contributions are non-universal because they depend on the lengths of the bonds. We analyzed length spectra and compared them to those deduced from the exact trace formula for quantum graphs in order to illustrate the dominance of these orbits in the corresponding spectra. Actually, their presence seems to be unavoidable and their effect on the spectral properties of the associated quantum graph is comparable to that of bouncing-ball orbits in a stadium billiard~\cite{Sieber1993}. However, in distinction to the latter, they occur in all possible realizations of periodic orbits with a certain period, i.e., their number is not of measure zero. In fact,  
it is impossible to extract the contributions of the shortest periodic orbits from the eigenvalue spectra by adding the slow oscillations, caused by them in the integrated spectral density (see Fig.~\ref{Fig61}), to the smooth part of the latter~\cite{Sieber1993}, which yields the unfolded eigenvalues. We determined these contributions by integrating the semiclassical trace formula Eq.~(\ref{trace}) and also from the inverse Fourier transform of $\tilde\rho(l)$, defined in Eq.~(\ref{Eq.1}) and obtained by using 1800 computed eigenvalues, where we took into account only the lengths intervals around peaks in the length spectra corresponding to periodic orbits of twice the lengths of the bonds~\cite{Dietz2005}; see Fig.~\ref{Fig6}. Both procedures yielded exactly the same results, thus corroborating the powerfulness of the former. However, even when including short primitive periodic orbits of period $n_p\leq 4$~\cite{Schanz2003}, which, e.g., in Fig.~\ref{Fig6} would comprise all peaks below $l\simeq 15$~dm, we were not able to achieve agreement with the GOE as concerns long-range fluctuations. Note, that backscattering, and thus the presence of shortest periodic orbits, may be avoided by introducing circulators in microwave networks, or considering unidirectional bonds in numerical simulations of quantum graphs. This, however, will lead to a violation of \T invariance~\cite{Hul2004,Lawniczak2008}. Furthermore, a considerably improved agreement with an extension of the GOE to spectra with missing levels may be achieved even in numerical simulations by simply randomly extracting eigenvalues from a level sequence. Then, small deviations are only visible in the domain of the statistical measures for long-range fluctuations, where the effects of non-universality set in. We corroborated these results in further experiments and in numerical simulations with a varying number of bonds and connectivity of the bonds, and in all cases observed similar deviations from the GOE predictions. 

We come to the conclusion that, due to the presence of shortest periodic orbits confined to individual bonds, a quantum graph might not provide a suitable model system for the investigation of long-range fluctuations in the level sequences of a generic, classically chaotic quantum system. This is corroborated by the deviations observed between the parameter-dependent spectral statistics of large graphs with Neumann boundary conditions at the vertices and the GOE predictions.~\cite{Hul2011}. Furthermore, it was shown there, that on the contrary, when replacing the vertex scattering matrices by matrices from the circular orthogonal ensemble, they are in good agreement with RMT predictions~\cite{Dietz1996}. We would like to emphasize that according to Refs.~\cite{Gnutzmann2004,Pluhar2013,Pluhar2013a,Pluhar2014} and experimental results presented, e.g., recently in Refs.~\cite{Rehemanjiang2016,Bialous2016} the introduction of time-reversal invariance violation or, similarly, unidirectionality, leads to a most appropriate system for the study of GSE or GUE features.     

B.D. thanks Tsampikos Kottos and Holger Schanz for fruitful discussions. This work was partially supported by the Ministry of Science and Higher Education (MSHE) Grant No. UMO-2013/09/D/ST2/03727 and the EAgLE international project (FP7-REGPOT-2013-1, Project No. 316014), co-financed in the years 2013-2016 by the MSHE.

\end{document}